\documentclass{mn2e}
\usepackage{graphicx}

\begin{document}

\title[The exterior gravitational field of rapidly rotating neutron 
stars]{Towards a self-consistent relativistic model of the exterior 
gravitational field of rapidly rotating neutron stars}

\author[Matthias Stute, Max Camenzind]{Matthias Stute,$^1$ Max Camenzind$^1$ \\
$^1$ Landessternwarte Heidelberg-K\"onigstuhl, D-69117 Heidelberg, 
Germany}

\maketitle

\begin{abstract}
We present a self-consistent, relativistic model of rapidly rotating 
neutron stars describing their exterior gravitational field. This is achieved 
by matching the new solution of Einstein's field equations found by Manko et 
al. \shortcite{MSM} and the numerical results for the interior of neutron 
stars with different equations of state calculated by Cook et al. 
\shortcite{CST}. This matching process gives constraints for the choice of the 
five parameters of the vacuum solution. Then we investigate some properties of 
the gravitational field of rapidly rotating neutron stars with these fixed 
parameters.
\end{abstract}

\section{Introduction}

Accretion on to neutron stars is primarily determined by their exterior 
gravitational field for which an analytical description is preferable to a 
numerical one. The solutions of Einstein's field equations suitable for Black 
Holes as the Schwarzschild- and the Kerr-metric do not hold for neutron stars 
because those have a solid surface and are therefore oblate when rapidly 
rotating. The second difficulty is the possible existence of an intrinsic 
magnetic field. A new solution has been strongly needed including the two 
parameters of the Kerr-metric, mass and angular momentum, and three additional 
ones, the mass quadrupole moment due to the oblateness of the star, the 
electric charge (which is of no importance for astrophysical objects) and the 
magnetic dipole moment. After more than ten years of work Manko et al. 
\shortcite{MSM} found a space-time meeting all criteria.

These parameters are in a mathematical sense formally independent, but in an
astrophysical sense, the mass quadrupole moment is a function of the mass and 
the angular momentum. To find these dependencies we use numerical results 
found by Cook et al. \shortcite{CST}. They solved the general relativistic 
structure equations for an axisymmetric rotating star for 14 different 
equations of state (EOS) and three sequences and presented certain quantities 
for the stars in a tabulated form for five of them. By comparing those with 
the analytical values we get correlations for the parameters of the new 
solution. 

This was also done by Sibgatullin \& Sunyaev \shortcite{SiS1}, but with 
another metric \cite{MMR} and only three EOS with low stiffness. These authors 
investigated their found models in further publications (the most recent is 
Sibgatullin \& Sunyaev \shortcite{SiS2}). This metric, however, did not permit 
the rational function representation -- the reason why Manko et al. continued 
their search for a new solution -- and could not be written in a concise form 
which is more useful for analytical examinations. Furthermore the same 
procedure -- but in opposite order --  was done for this new metric extended 
to all five EOS and all three sequences.

With fixed parameters this new solution is highly useful to improve the models
of rapidly rotating neutron stars e.g. in accretion disc calculations for which
the Schwarzschild- or Kerr-metric are still used at this time. These introduce 
large errors because of their significant deviations to real neutron stars due 
to symmetry simplifications which are now not necessary anymore.

\section{The general exterior field of rotating neutron stars} \label{ext}

Starting with the most general form of a stationary axi\-symmetric metric, 
found by Papapetrou \shortcite{Pap}, 
\begin{equation} \label{Metrik}
ds^2 = f \, (dt - w \, d\phi)^2 - f^{-1} [ \, e^{2 \gamma} \, (d\rho^2 +
dz^2) + \rho^2 \, d\phi^2 ] ,
\end{equation}
where $\rho$ and $z$ are quasi-cylindrical Weyl-Papapetrou-coordinates and 
$f$, $w$ and $\gamma$ are functions independent of $\phi$ and $t$, 
Ernst \shortcite{Ern} found that Einstein's equations reduce to the form
\begin{eqnarray}
( Re \, \mathcal{E} + |\Phi|^2 ) \nabla^2 \mathcal{E} &=&
(\nabla \mathcal{E} + 2 \Phi^{*} \nabla \Phi) \cdot \nabla \mathcal{E} 
\nonumber \\ \label{Ernst2}
( Re \, \mathcal{E} + |\Phi|^2 ) \nabla^2 \Phi &=&
(\nabla \mathcal{E} + 2 \Phi^{*} \nabla \Phi) \cdot \nabla \Phi,
\end{eqnarray}
after introducing the two complex potentials $\mathcal{E}$ and $\Phi$ and their
complex conjugates $\mathcal{E}^{*}$ and $\Phi^{*}$ as
\begin{equation} \label{Ernstpot}
\mathcal{E} = ( f - |\Phi|^2 ) + i \varphi \qquad
\Phi = A_{t} + i \varphi' ,
\end{equation}
$\varphi$ and $\varphi'$ are defined by 
\begin{eqnarray} \label{varphi}
\rho^{-1} f^2 \nabla w - 2 \hat n \times Im(\Phi^{*} \nabla \Phi)
&=& \hat n \times \nabla \varphi \nonumber \\
\rho^{-1} f (\nabla A_{\phi} - w \nabla A_{t}) &=& \hat n \times \nabla 
\varphi'
\end{eqnarray} 
with the $t$- and $\phi$-components $A_{t}$ and $A_{\phi}$ of the 
electromagnetic four-potential and the unit-vector in azimuthal direction 
$\hat n$. If the Ernst-potentials are known, the functions $f$, $w$ and 
$\gamma$ can be directly calculated from 
\begin{eqnarray} \label{metricfunctions}
f &=& Re \mathcal{E} + \Phi \Phi^{*} \nonumber \\
w_{,\rho} &=& - \rho f^{-2} [ \varphi_{,z} + 2 \, Im(\Phi^{*} \Phi_{,z}) ]
\nonumber \\
w_{,z} &=& \rho f^{-2}  [ \varphi_{,\rho} + 2 \, Im(\Phi^{*} \Phi_{,\rho}) ]
\nonumber \\
\gamma_{,\rho} &=& \frac{1}{4} \rho f^{-2} [ (\mathcal{E}_{,\rho} + 2 \, 
\Phi^{*} \Phi_{,\rho}) (\mathcal{E}_{,\rho}^{*} + 2 \, \Phi \Phi_{,\rho}^{*}) 
\nonumber \\
&& - (\mathcal{E}_{,z} + 2 \, \Phi^{*} \Phi_{,z}) (\mathcal{E}_{,z}^{*} + 2 \, 
\Phi  \Phi_{,z}^{*}) ] \nonumber \\
&& - \rho f^{-1} (\Phi_{,\rho} \Phi_{,\rho}^{*} - 
\Phi_{,z} \Phi_{,z}^{*})
\nonumber \\
\gamma_{,z} &=& \frac{1}{2} \rho f^{-2} Re[ (\mathcal{E}_{,\rho} + 2 \, 
\Phi^{*} \Phi_{,\rho}) (\mathcal{E}_{,z}^{*} + 2 \, \Phi
\Phi_{,z}^{*}) ] 
\nonumber \\
&& - 2 \rho f^{-1} Re(\Phi_{,\rho}^{*} \Phi_{,z}) .
\end{eqnarray}
The two Ernst-equations are solved by simple integral equations 
\cite{Sib} which introduce the axis data of the Ernst-potentials $e(z) = 
\mathcal{E} (\rho = 0,z)$ and $f(z) = \Phi (\rho = 0,z)$: 
\begin{equation} \label{integralsolution}
\mathcal{E} = \frac{1}{\pi} \int \limits^{1}_{-1} \frac{e(\xi) \mu_{1} 
d\sigma}{\sqrt{1-\sigma^2}} \qquad
\Phi = \frac{1}{\pi} \int \limits^{1}_{-1} \frac{f(\xi) \mu_{1} 
d\sigma}{\sqrt{1-\sigma^2}} ,
\end{equation}
where $\mu_{1}$ has to satisfy
\begin{eqnarray} \label{constraints_mu}
\int \limits^{1}_{-1} \! \! \! \! \! \! \! \! \; - \, \, \,
 \frac{\mu_{1} (\sigma) \, ( \tilde e(\eta) + e(\xi)
+ 2 \tilde f(\eta) f(\xi) ) \, d\sigma}{(\xi - \eta) \,
\sqrt{1-\sigma^2}} &=& 0 \nonumber \\
\int \limits^{1}_{-1} \frac{\mu_{1} (\sigma) \,
d\sigma}{\sqrt{1-\sigma^2}} &=& \pi 
\end{eqnarray}
with $\eta = z + i \rho \tau$, $\xi = z + i \rho \sigma$, $\tilde e(\xi) = 
\bar e(\bar \xi)$, $\tau,  \sigma \in [-1,1]$. In the first integral the Cauchy
principal value has to be calculated. These new functions can be expressed as 
a multipole expansion of the mass-, angular momentum-, electric and magnetic 
field-distribution 
\begin{equation} \label{multipole1}
\mathcal{E} = \frac{1 - \xi}{1 + \xi} \qquad \Phi = \frac{q}{1 + \xi}
\end{equation}
\begin{eqnarray} \label{multipole2}
\xi (\rho = 0) &=& \sum^{\infty}_{n = 0} m_{n} z^{-(n+1)} \nonumber \\
q (\rho = 0) &=& \sum^{\infty}_{n = 0} q_{n} z^{-(n+1)} .
\end{eqnarray}
The physical multipole moments are the real (mass multipoles) and imaginary 
(angular momentum multipoles) part of $m_{n}$ and the real (electric 
multipoles) and imaginary (magnetic  multipoles) part of $q_{n}$.

To derive a new exact analytical solution of Einstein's equations one starts 
with an ansatz of $e(z)$ and $f(z)$, checks whether the solution has the 
wanted parametrisation of the multipole moments and performs the cumbersome but
straight-forward calculation of the metric functions $f$, $w$ and $\gamma$.

Different authors (e.g. Aguirregabaria et al. \shortcite{ACM}, Manko et al. 
\shortcite{MMR}) made an effort to find a new solution involving five 
parameters representing mass, angular momentum, charge, magnetic dipole moment 
and mass quadrupole moment of the star which can be expressed with 
relatively simple rational functions. After more than ten years and several 
ansatzes such a solutions was found \cite{MSM} starting 
with
\begin{eqnarray} \label{Manko_axisdata}
e(z) &=& \frac{(z - m - i a) \, (z + i b) + d - \delta - a b}{(z + m - i
a) \, (z + i b) + d - \delta - a b} \nonumber \\
f(z) &=& \frac{Q z + i \mu}{(z + m - i a) \, (z + i b) + d - \delta - a
b}
\end{eqnarray}
with
\begin{eqnarray}
\delta &=& \frac{\mu^2 - m^2 b^2}{m^2 - (a - b)^2 - Q^2} \nonumber \\
d &=& \frac{1}{4} \, [ m^2 - (a - b)^2 - Q^2 ] .
\end{eqnarray}
After checking the multipole moments one finds that $m$ is the mass, $a$ the 
specific angular momentum, $Q$ the charge and $\mu$ the magnetic dipole moment
(if $Q = 0$). The parameter $b$ is no physical quantity, but is related to 
the mass quadrupole moment. For comparison the functions $e(z)$ and $f(z)$ of 
the Schwarzschild- and the Kerr-solution are
\begin{equation} \label{Schwarzschild_axisdata}
e(z) = \frac{z - m}{z + m} = 1 - \frac{2 m}{z + m} \qquad f(z) = 0 
\end{equation}
\begin{equation}\label{Kerr_axisdata}
e(z) = \frac{z - m + i\,a}{z + m + i\,a} \qquad f(z) = 0 .
\end{equation}
To be able to write the metric function in a rational form it is convenient to 
introduce general spheroidal coordinates by
\begin{equation} \label{spheroidalcoord}
x = \frac{1}{2 \kappa} (r_{+} + r_{-}) \qquad y = \frac{1}{2 \kappa} (r_{+} 
- r_{-})
\end{equation}
with $r_{\pm} = \sqrt{\rho^2 + (z \pm \kappa)^2}$ and $\kappa = \sqrt{d 
+ \delta}$. The metric then becomes 
\begin{eqnarray} \label{metric_spheroidal}
ds^2 &=& f \, (dt - w \, d\phi)^2 - \kappa^2 f^{-1} \Bigl[ e^{2 \gamma} \,
(x^2 - y^2) \Bigl( \frac{dx^2}{x^2 - 1} \nonumber \\
&& + \frac{dy^2}{1 - y^2} \Bigr) + (x^2 - 1) (1 - y^2) d\phi^2 \Bigr]
\end{eqnarray}
and the functions are
\begin{eqnarray} \label{metricfunctions_manko}
f = \frac{E}{D} \qquad e^{2 \gamma} = \frac{E}{16 \kappa^8 (x^2 -
y^2)^4} \qquad w = \frac{- (1 - y^2) F}{E} \nonumber \\
\end{eqnarray}
with
\begin{eqnarray} \label{metricfunctions_abbr}
E &=& \{ 4 [ \kappa^2 (x^2 - 1) + \delta (1 - y^2) ]^2 + (a - b) \nonumber \\
&& [ (a - b) (d - \delta) - m^2 b + Q \mu ] (1 - y^2)^2 \}^2  \nonumber \\
&& - 16 \kappa^2 (x^2 - 1) (1 - y^2) \{ (a - b) [ \kappa^2 (x^2 - y^2) 
\nonumber \\
&& + 2 \delta y^2 ] + (m^2 b - Q \mu) y^2 \}^2 \nonumber \\
D &=& \{ 4 (\kappa^2 x^2 - \delta y^2)^2 + 2 \kappa m x [ 2 \kappa^2
(x^2 - 1) \nonumber \\
&& + (2 \delta + a b - b^2) (1 - y^2) ] + (a - b) [ (a - b) (d - \delta) 
\nonumber \\
&& - m^2 b + Q \mu] (y^4 - 1) - 4 d^2 \}^2 + 4 y^2 \{ 2 \kappa^2 (x^2 - 1) 
 \nonumber \\
&& [ \kappa x (a - b) - m b ] - 2 m b \delta (1 - y^2)+ [ (a - b) \nonumber \\
&& (d - \delta) - m^2 b + Q \mu] (2 \kappa x + m) (1 - y^2) \}^2 \nonumber \\
F &=& 8 \kappa^2 (x^2 - 1) \{ (a - b) [ \kappa^2 (x^2 - y^2) + 2
\delta y^2 ] \nonumber \\
&& + y^2 (m^2 b - Q \mu) \} \{ \kappa m x [ (2 \kappa x + m)^2 
- 2 y^2 (2 \delta + a b  \nonumber \\
&& - b^2) - a^2 + b^2 - Q^2 ] - 2 \kappa^2 Q^2 x^2 - 2 y^2 (4 \delta d 
\nonumber \\
&& - m^2 b^2) \} + \{ 4 [ \kappa^2 (x^2 - 1) + \delta (1 - y^2) ]^2 
\nonumber \\
&& + (a - b) [ (a - b) (d - \delta) - m^2 b + Q \mu ]  \nonumber \\
&& (1 - y^2)^2 \} (4 (2 \kappa m b x + 2 m^2 b - Q \mu)  \nonumber \\
&& [ \kappa^2 (x^2 - 1) + \delta (1 - y^2) ] + (1 - y^2)  \nonumber \\
&& \{ (a - b) (m^2 b^2- 4 \delta d) - (4 \kappa m x + 2 m^2  \nonumber \\
&& - Q^2) [ (a - b) (d - \delta) - m^2 b + Q \mu] \} ) .
\end{eqnarray}

Mathematically, the five parameters of this solution are independent. To make 
it applicable for astrophysical purposes and to describe astrophysical 
configurations, one needs to find constraints for the choice of and the 
relations between some of these parameters. Those can only be found by 
matching the exterior solution with existing models for the interior of 
neutron stars.

\section{The Interior of neutron stars} \label{int}

Until now, the knowledge of the composition of the interior of neutron 
stars is very poor. A large zoo of different EOS exists to describe all 
possible ingredients of the star and the interactions of these particles. 
Before the right EOS can be singled out by observations, i.e. determining the 
mass-radius-relation by e.g. measuring the cooling curve of neutron stars, the 
basic structure equations for rotating, axisymmetric stars -- similar to the 
Oppenheimer-Volkoff-equations for spherical stars -- have to be solved 
numerically for each model. Cook et al. \shortcite{CST} performed these 
calculations for 14 different EOS with a modified variation of the KEH-Code 
\cite{KEH} and tabulated several quantities for five of them representing the 
whole range of stiffness (A, AU, FPS, L, M; see table \ref{tableEOS}, Cook et 
al. \shortcite{CST} and references therein). For further informations on 
different numerical methods see e.g. Ansorg et al. \shortcite{AKM} and 
references therein.
\begin{table}
\caption{Equations of state, used in Cook et al. \shortcite{CST}}
\label{tableEOS}
\begin{tabular}{ll}
\hline
EOS & Description \\
\hline\hline
A & Reid soft core, adapted to nuclear matter \\
AU & Argonne V14 + Urbana VII \\
FPS & Urbana V14 + Three Nuclei Interaction \\
L & Nuclear attraction due to scalar exchange \\
M & Nuclear attraction due to pion exchange \\
& (Tensor interaction) \\
\hline
\end{tabular}
\end{table}
For each EOS these authors computed models for three star sequences: first the 
normal sequence (NS) of stars whose masses are below the maximum mass 
of non-rotating stars, second the maximum mass normal sequence (MM) of stars 
with the maximum mass of non-rotating stars and third the supramassive 
sequence (SM) of stars whose masses exceed the maximum mass of non-rotating 
stars. These numerical values can be compared with the analytically 
derived ones to adjust the free parameters. The following fitting procedure 
was performed for all five equations of state and all three sequences.

There exists a newer EOS including more effects of nuclear and particle 
physics (Glendenning \& Weber \shortcite{GlW} and references therein), but 
unfortunately the analog analysis of Cook et al. \shortcite{CST} has not yet 
been done. Perhaps this modern EOS describes the interior of neutron stars in 
a better way than the older EOS considered by Cook et al., then the same 
fitting procedure performed with this EOS would give more physically 
meaningful results. More observations are needed to clear this point.

\section{Fitting the free parameters} \label{fit}

\subsection{The gravitational redshift}

One of the quantities tabulated in Cook et al. \shortcite{CST} is the 
gravitational redshift photons experience being emitted forward ($Z_{f}$) and 
backward ($Z_{b}$) with respect to the rotation direction at the equator. The 
same can be easily calculated for this analytical solution. 
For photons which are emitted at the equator in forward (backward) direction 
with respect to the rotational direction the four-momentum has the form 
\begin{equation}
p^{\mu} = \textrm{const.} \, \times \,[ \, \xi_{(t)}^{\mu} +
\frac{f}{f w \pm \rho} \, \xi_{(\phi)}^{\mu} \, ]
\end{equation}
with the Killing-vectors $\xi_{(t)}^{\mu}$ and $\xi_{(\phi)}^{\mu}$. The 
frequency is calculated by
\begin{equation}
\omega_{E} = p_{\mu} \, u^{\mu},
\end{equation}
where the fluid four-velocity on the equator is
\begin{equation}
u^{\mu} = \frac{f^{-1/2}}{\sqrt{1 - v^2}} \, ( \xi_{(t)}^{\mu} + \Omega
\, \xi_{(\phi)}^{\mu} )
\end{equation}
and the fluid velocity measured by a ZAMO (zero angular momentum observer)  
\begin{equation}
v = \sqrt{2 w \Omega - \Omega^2 \, (w^2 - \rho^2/f^2)} .
\end{equation}
The frequency observed at infinity is 
\begin{equation}
\omega_{\infty} = p_{\mu} \, \xi_{(t)}^{\mu} .
\end{equation}
Therefore the redshift which is looked for has the form
\begin{eqnarray}
Z_{^f_b} &=& \frac{p_{\mu} \, u^{\mu}}{p_{\mu} \, \xi_{(t)}^{\mu}} - 1
\nonumber \\
&=& \left( \pm 2 f \rho \sqrt{1 - v^2} - 2 f^{3/2} (-1 + v^2 ) (\mp \rho + f
w)  \right. \nonumber \\
&& \left. + \sqrt{1 - v^2} \sqrt{4 f^2 \rho^2  v^2  - 4 f^4  (-1 + v^2 ) w^2
} \right)/ \nonumber \\
&& \left( 2 f^{3/2} (-1 + v^2 ) (\mp \rho + f w) \right) .
\end{eqnarray}
A suitable combination of these two redshifts is
\begin{equation}
Z := (1 + Z_{f}) \, (1 + Z_{b}) = \frac{1}{f (\rho_{eq.}, z_{eq.})} 
\end{equation}
with the metric function $f$ taken at the equator of the neutron star. To be 
able to compare the analytical function
\begin{equation}
Z_{anal.} = Z_{anal.} (\rho, z = 0; m, a, b, Q, \mu)
\end{equation}
with the numerical results one has to match the parameters. Cook et al. 
\shortcite{CST} considered only neutron star models without electric 
charge and magnetic field, therefore one can set
\begin{equation}
Q = 0 \qquad \textrm{and} \qquad \mu = 0 .
\end{equation}
Further it is useful to introduce the specific angular momentum $j$ 
defined by
\begin{equation}
j = \frac{J}{m^2} = \frac{a}{m} \qquad \Longrightarrow \qquad a = m \, j .
\end{equation}
Because of the fact that Cook et al. \shortcite{CST} made their calculations 
in Boyer-Lindquist-like quasi-spherical coordinates $r$ and $\theta$ one has 
to transform the coordinates as 
\begin{eqnarray} \label{coord_transform}
r &=& \frac{1}{2} \, \Bigl(\sqrt{\rho^2 + (z + \kappa)^2} + \sqrt{\rho^2 + (z - 
\kappa)^2} \Bigr) + m \nonumber \\
\theta &=& \arccos{\frac{1}{2\,\kappa} \, \Bigl(\sqrt{\rho^2 + (z + 
\kappa)^2} - \sqrt{\rho^2 + (z - \kappa)^2} \Bigr)}
\end{eqnarray}
The connection of the distance coordinates in the equatorial plane ($z = 0$) 
is then
\begin{eqnarray}
r_{*} &=& \frac{R_{*}}{m} = \frac{1}{m} \, \left( m +
\sqrt{\rho_{*}^2 + \kappa^2} \right) \nonumber \\
\rho_{*} &=& \sqrt{m^2 \, (r_{*}-1)^2 - \kappa^2} 
\end{eqnarray}
with the equatorial stellar radius $R_{*}$. Combining all these equations one 
gets 
\begin{equation}
b = b [ Z (j), r (j); m (j), j ] = b (j) .
\end{equation}
It follows from the numerical data that
\begin{eqnarray}
M_{*} &=& M_{*} (j) = M_{*} (-j) \nonumber \\
R_{*} &=& R_{*} (j) = R_{*} (-j) \nonumber \\
Z_{num.} &=& Z_{num.} (j) = Z_{num.} (-j) 
\end{eqnarray}
so $b (j)$ also has to be symmetric with respect to the rotation direction. As 
we wanted to fit the parameter $b$ in the whole range [-1,1] for $j$ we had to 
include the Schwarzschild limiting case for $j = 0$. The solution of Manko et 
al. \shortcite{MSM} reduces to Kerr by setting
\begin{equation} \label{bKerr}
b = \sqrt{a^2 - m^2}
\end{equation}
and to Schwarzschild with $b = i m$. Therefore, we were especially interested 
in imaginary values of $b$. This poses no problem as $b$ is no physical 
observable. Moreover, it can be shown that the equation of the 
marginally-stable orbit (next section) has solutions for small $j$ only with 
imaginary $b$.

\subsection{The marginally-stable orbit -- an independent test}

The radius of the marginally-stable orbit is another quantity given in the 
tables of Cook et al. \shortcite{CST}. This provides the 
possibility to perform an independent test of the fits done above. The 
following calculation gives the analytic expression for this radius. 
Consider a particle moving in the equatorial plane. With the three 
constants of motion $\mathbf{E}$, $\mathbf{L}$ and $\mathbf{K} = 1$ related to 
the energy, the angular momentum and the mass of the particle the equations of 
motion are fully integrable. The equation of the radial coordinate $\rho$ can 
be interpreted as energy equation with an effective potential 
\begin{equation}
V(\rho) = \frac{e^{2 \gamma}}{f} \dot \rho^2 = \frac{\mathbf{E}^2}{f}
- \frac{f}{\rho^2} \, (\mathbf{L} - \mathbf{E} w)^2 - 1 .
\end{equation}
For circular orbits the conditions 
\begin{eqnarray}
V(\rho) = \frac{\mathbf{E}^2}{f} - \frac{f}{\rho^2} \, (\mathbf{L} -
\mathbf{E} w)^2 - 1 &=& 0 \\
\frac{dV(\rho)}{d\rho} = \frac{2\,f\,( \mathbf{L} - \mathbf{E}\,w )^2}{\rho^3}
- \frac{( \mathbf{L} - \mathbf{E}\,w
)^2\,f_{,\rho}}{\rho^2}  && \nonumber \\
- \frac{\mathbf{E}^2\,f_{,\rho}}{f^2} + \frac{2\,\mathbf{E}\,f\,( \mathbf{L} -
\mathbf{E}\,w ) \,w_{,\rho}}{\rho^2} &=& 0
\end{eqnarray}
have to hold from which one can calculate
\begin{eqnarray} \label{msbE}
\mathbf{E} &=& \frac{\sqrt{f}}{\sqrt{1 -
f^2\,\mathcal{X}^2/\rho^2}} \\
\mathbf{L} &=& \mathbf{E} \, (w + \mathcal{X}) \label{msbL} \\
\mathcal{X} &=& \frac{\rho\,[ - w_{,\rho}\,f^2
- \sqrt{w_{,\rho}^2\,f^4 + f_{,\rho}\,\rho\,( 2\,f - f_{,\rho}\,\rho
)}\, ]}{f\,( 2\,f - f_{,\rho}\,\rho )} . \label{msbX}
\end{eqnarray}
With the condition for a marginally-stable circular orbit
\begin{eqnarray}
\frac{d^2V(\rho)}{d\rho^2} = \frac{-6\,f\,( \mathbf{L} - \mathbf{E}\,w
)^2}{\rho^4} + \frac{4\,( \mathbf{L} - \mathbf{E}\,w
)^2\,f_{,\rho}}{\rho^3}  && \nonumber \\
- \frac{8\,\mathbf{E}\,f\,( \mathbf{L} - \mathbf{E}\,w )\,w_{,\rho}}{\rho^3} 
+ \frac{4\,\mathbf{E}\,( \mathbf{L} - \mathbf{E}\,w )\,
f_{,\rho}\,w_{,\rho}}{\rho^2}  && \nonumber \\
+ \frac{2\,\mathbf{E}^2\,f_{,\rho}^2}{f^3} -
\frac{2\,\mathbf{E}^2\,f\,w_{,\rho}^2}{\rho^2} 
- \frac{\mathbf{E}^2\,f_{,\rho\rho}}{f^2}  && \nonumber \\
- \frac{( \mathbf{L} -
\mathbf{E}\,w )^2\,f_{,\rho\rho}}{\rho^2} + \frac{2\,\mathbf{E}\,f\,(
\mathbf{L} - \mathbf{E}\,w )\,w_{,\rho\rho}}{\rho^2} &=& 0 \nonumber \\
\end{eqnarray}
and with (\ref{msbE}), (\ref{msbL}) and (\ref{msbX}) the equation for the 
radius of the marginally-stable orbit is
\begin{eqnarray} \label{msB-Gl}
\Bigl( w_{,\rho}\,w_{,\rho\rho}\,f^5\,\rho\,(2 f - f_{,\rho}\,\rho) + 
w_{,\rho}^2\,f^4\,[ 2 f^2 + ( -f_{,\rho}^2  && \nonumber \\
+ f_{,\rho\rho}\,f)\,\rho^2] + w_{,\rho}\,f^2\,\sqrt{w_{,\rho}^2\,f^4 + 
f_{,\rho}\,\rho\,(2 f - f_{,\rho}\,\rho)} && \nonumber \\
\,[ 2 f^2 
+ 2\,f_{,\rho}^2\,\rho^2 - f\,\rho\,(4 f_{,\rho} + f_{,\rho\rho}\,\rho) ] 
+ \rho\,(2 f - f_{,\rho}\,\rho) && \nonumber \\
\,\{ 3 f_{,\rho}\,f^2 
- 4 f_{,\rho}^2\,f\,\rho + f_{,\rho}^3\,\rho^2 + f^2\,[ f_{,\rho\rho}\,\rho 
 && \nonumber \\
- w_{,\rho\rho}\,f\,\sqrt{w_{,\rho}^2\,f^4 + f_{,\rho}\,\rho\,(2 f -
f_{,\rho}\,\rho)} ] \} \Bigr) / && \nonumber \\
\Bigl( f^2\,\rho^2\,\{ w_{,\rho}^2\,f^4 + 3
f_{,\rho}\,f\,\rho - f_{,\rho}^2\,\rho^2  && \nonumber \\
- f^2\,[ 2 + w_{,\rho}\,\sqrt{w_{,\rho}^2\,f^4 + 
f_{,\rho}\,\rho\,(2 f - f_{,\rho}\,\rho)}] \} \Bigr) &=& 0 . \nonumber \\
\end{eqnarray}

Again, one has to set $z = Q = \mu = 0$, then a function of $m$, 
$r$, $j$ and $b$ remains. To test the fitted parameters we used the $m$- and 
$b$-values of the tables \ref{mfit} and \ref{Z_bfit} and made implicit plots 
in the ($r$,$j$)-plane. 

\section{The fits and the tests}

\begin{figure}
\resizebox{\hsize}{!}{\includegraphics{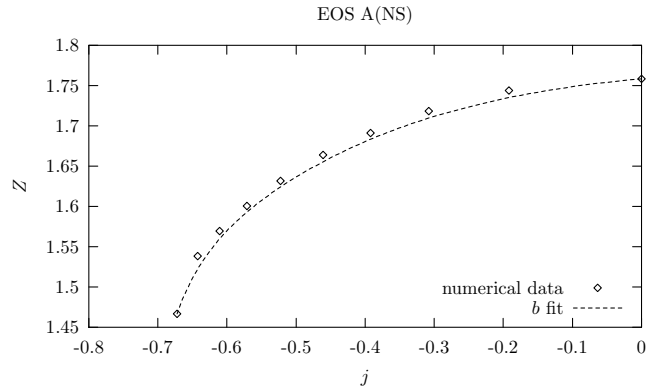}}
\caption{gravitational redshift $Z$ vs. rotation parameter $j$ for the normal 
sequence of EOS A}
\label{Z_ANS}
\end{figure}

\begin{figure}
\resizebox{\hsize}{!}{\includegraphics{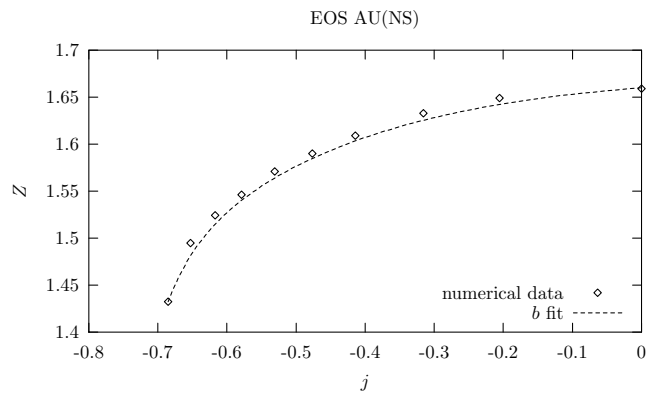}}
\caption{gravitational redshift $Z$ vs. rotation parameter $j$ for the normal 
sequence of EOS AU}
\label{Z_AUNS}
\end{figure}

\begin{figure}
\resizebox{\hsize}{!}{\includegraphics{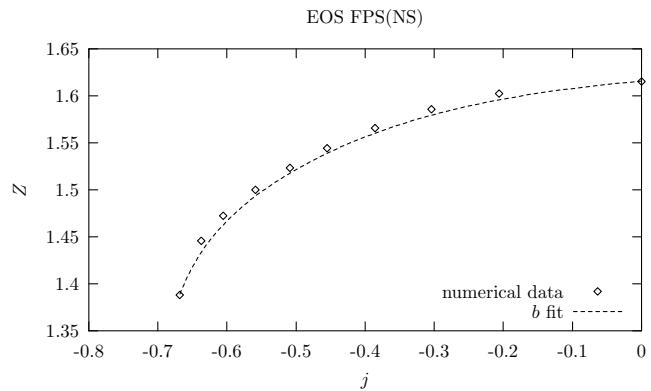}}
\caption{gravitational redshift $Z$ vs. rotation parameter $j$ for the normal 
sequence of EOS FPS}
\label{Z_FPSNS}
\end{figure}

\begin{figure}
\resizebox{\hsize}{!}{\includegraphics{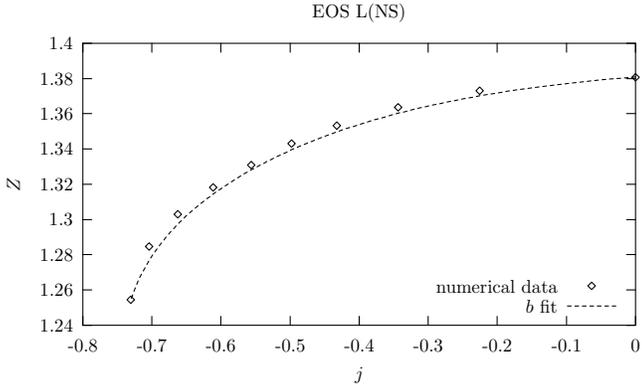}}
\caption{gravitational redshift $Z$ vs. rotation parameter $j$ for the normal 
sequence of EOS L}
\label{Z_LNS}
\end{figure}

\begin{figure}
\resizebox{\hsize}{!}{\includegraphics{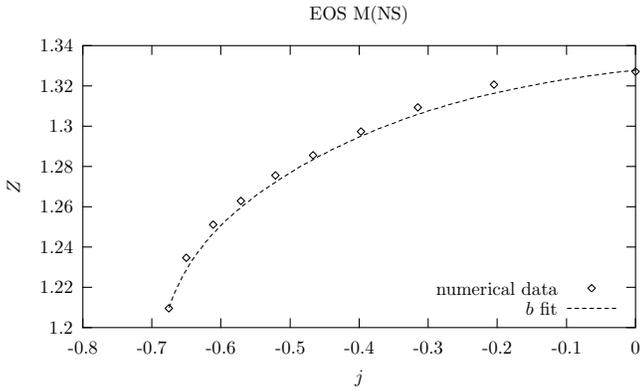}}
\caption{gravitational redshift $Z$ vs. rotation parameter $j$ for the normal 
sequence of EOS M}
\label{Z_MNS}
\end{figure}

Fitting the numerically calculated masses in Cook et al. \shortcite{CST} with
\begin{eqnarray} \label{m_ansatz}
m(j) &=& m_{0} + m_{1} j^2 + m_{2} j^4 + m_{3} j^6 
\end{eqnarray}
and choosing the ansatz
\begin{equation} \label{b_ansatz}
b = \sqrt{\Delta_{1}\,a^2 - m^2 + \Delta_{2}\,\frac{a^4}{m^2} +
\Delta_{3}\,\frac{a^6}{m^4}}
\end{equation}
on finds the fitting parameters as listed in table \ref{Z_bfit} and table 
\ref{mfit}. This ansatz is chosen to ensure $b = i m$ for $j = 0$, to 
achieve $b (j) = b (-j)$ and to be sure that all terms have the right units. 
The fit parameters in (\ref{m_ansatz}) are in units of solar masses and the 
those in (\ref{b_ansatz}) are dimensionless. Although $b$ is imaginary the 
parameter $\kappa$ stays pure real. Therefore this solution is subextreme.  
Pure imaginary values of $\kappa$ represent hyperextreme spacetimes which 
could be interpreted as generated by relativistic discs.
\begin{table}
\caption{Dimensionless fitting parameters of the fit ansatz 
\mbox{$b = \sqrt{\Delta_{1}\,a^2 - m^2 + \Delta_{2}\,\frac{a^4}{m^2} +
\Delta_{3}\,\frac{a^6}{m^4}}$}}
\label{Z_bfit}
\begin{tabular}{llll}
\hline
EOS (Seq.) & $\Delta_{1}$ & $\Delta_{2}$ & $\Delta_{3}$ \\
\hline\hline
A(NS) & $-7.97743$ & $-0.09978$ & $-2.62792$ \\
A(MM) & $-2.98207$ & $-12.02850$ & $14.27300$ \\
A(SM) & $0.51654$ & $-25.66350$ & $30.50040$ \\
AU(NS) & $-8.35758$ & $-7.59210$ & $11.33570$ \\
AU(MM) & $-2.10796$ & $-7.94790$ & $8.71811$ \\
AU(SM) & $0.83801$ & $-18.47420$ & $19.58550$ \\
FPS(NS) & $-8.06897$ & $-20.34330$ & $40.12250$ \\
FPS(MM) & $-3.94624$ & $-6.15936$ & $3.74551$ \\
FPS(SM) & $-0.44467$ & $-21.45520$ & $25.25300$ \\
L(NS) & $-19.75040$ & $6.10370$ & $-6.68697$ \\
L(MM) & $-3.59342$ & $-7.91550$ & $7.25291$ \\
L(SM) & $-0.10791$ & $-20.44890$ & $22.08320$ \\
M(NS) & $-23.07410$ & $20.14150$ & $-37.30740$ \\
M(MM) & $-6.50839$ & $-47.76620$ & $87.45900$ \\
M(SM) & $11.14010$ & $-144.35100$ & $221.00000$ \\
\hline
Kerr & $1$ & $0$ & $0$ \\
\hline
\end{tabular}
\end{table}
\begin{table}
\caption{Fitting parameters in units of solar masses of the fit ansatz 
$m(j) = m_{0} + m_{1} j^2 + m_{2} j^4 + m_{3} j^6 $} \label{mfit}
\begin{tabular}{lllll}
\hline
EOS (Seq.) & $m_{0}$ & $m_{1}$ & $m_{2}$ & $m_{3}$ \\
\hline\hline
A(NS) & $1.39999$ & $0.08294$ & $-0.01655$ & $-0.00135$ \\
A(MM) & $1.65510$ & $0.14606$ & $-0.02487$ & $-0.00866$ \\
A(SM) & $1.73043$ & $0.20082$ & $-0.08822$ & $0.03529$ \\
AU(NS) & $1.40004$ & $0.07455$ & $-0.01832$ & $0.00962$ \\
AU(MM) & $2.13344$ & $0.21933$ & $0.03005$ & $-0.03461$ \\
AU(SM) & $2.21414$ & $0.32712$ & $-0.12039$ & $0.07834$ \\
FPS(NS) &  $1.40003$ & $0.07255$ & $-0.02813$ & $0.02114$ \\
FPS(MM) & $1.79945$ & $0.16119$ & $-0.04416$ & $0.01113$ \\
FPS(SM) & $1.87636$ & $0.21989$ & $-0.11459$ & $0.05728$  \\
L(NS) &  $1.40000$ & $0.04002$ & $-0.00943$ & $0.00292$ \\
L(MM) & $2.69993$ & $0.23158$ & $0.00001$ & $-0.03842$ \\
L(SM) & $2.84355$ & $0.36050$ & $-0.19396$ & $0.11556$ \\
M(NS) & $1.39999$ & $0.03695$ & $-0.01482$ & $0.00563$ \\
M(MM) & $1.80458$ & $0.14357$ & $-0.27871$ & $0.31632$ \\
M(SM) & $1.87629$ & $0.38213$ & $-0.91656$ & $0.89839$ \\
\hline
\end{tabular}
\end{table}
The numerical data fall on the theoretical curves with an accuracy of 
\mbox{0.6 - 1.8 \%} (see figures \ref{Z_ANS}-\ref{Z_MNS}). The small deviation 
is a consequence of rounding and numerical errors during the two fit processes 
of the mass and of $b$ which could not be improved by taking higher orders 
into account. Because of the symmetry $Z (j) = Z (-j)$ only the 
negative $j$-values are plotted. Although this procedure was done for all 
sequences, we show only plots of the normal sequences, but the error range is 
valid for all sequences.

\begin{figure}
\resizebox{\hsize}{!}{\includegraphics{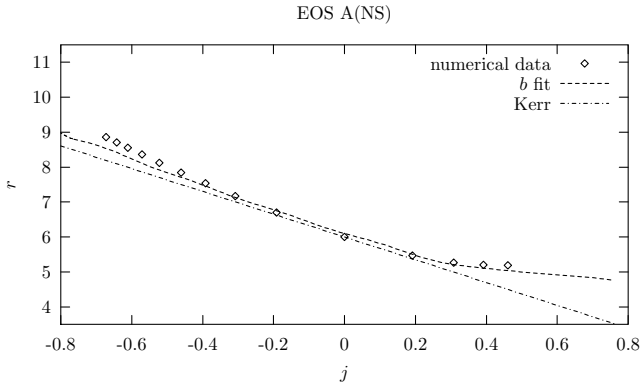}}
\caption{radius of the marginally-stable orbit (in units of $m$) vs. rotation 
parameter $j$ for the normal sequence of EOS A}
\label{curve_ANS}
\end{figure}

\begin{figure}
\resizebox{\hsize}{!}{\includegraphics{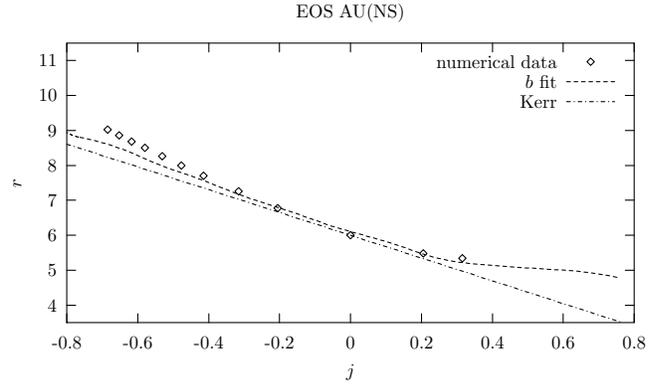}}
\caption{radius of the marginally-stable orbit (in units of $m$) vs. rotation 
parameter $j$ for the normal sequence of EOS AU}
\label{curve_AUNS}
\end{figure}

\begin{figure}
\resizebox{\hsize}{!}{\includegraphics{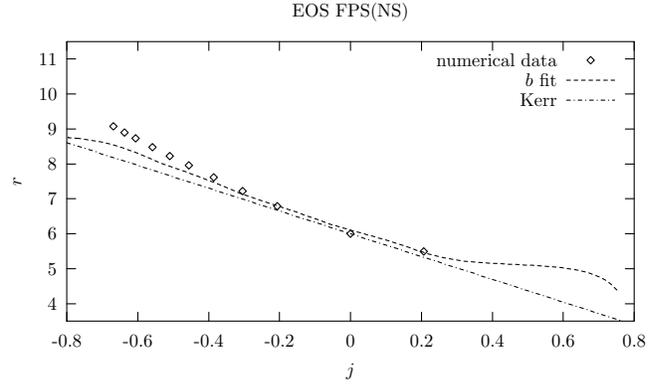}}
\caption{radius of the marginally-stable orbit (in units of $m$) vs. rotation 
parameter $j$ for the normal sequence of EOS FPS}
\label{curve_FPSNS}
\end{figure}

\begin{figure}
\resizebox{\hsize}{!}{\includegraphics{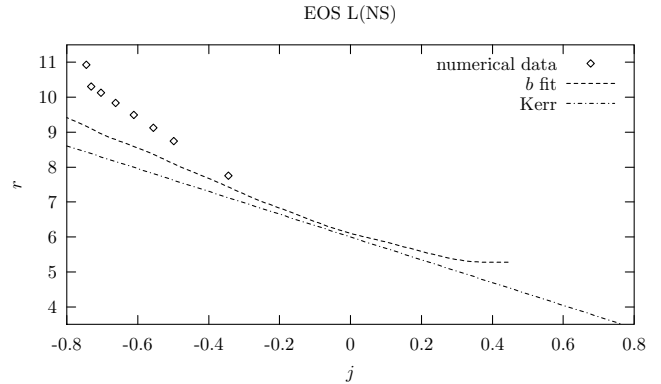}}
\caption{radius of the marginally-stable orbit (in units of $m$) vs. rotation 
parameter $j$ for the normal sequence of EOS L}
\label{curve_LNS}
\end{figure}

The test of these parameters with the radius of the marginally-stable orbit 
shows deviations between the numerical data and the analytical function of the 
order \mbox{4 - 8 \%}, in the case of the normal sequence of the \mbox{EOS L} 
15 \% (Figs. \ref{curve_ANS}-\ref{curve_LNS}). In these figures the same 
function is plotted for the Kerr-metric \cite{MiL} for 
comparison. The radius of the 
marginally-stable orbit of the normal sequence of the EOS M is always smaller 
than the stellar radius, so the corresponding plot is omitted.

\section{Symmetry with respect to the equatorial plane}

The ansatz of an imaginary parameter $b$ was a direct consequence of the 
requirement that Schwarzschild is a limiting case for vanishing rotation. This 
choice causes a problem. Following the formalism of deriving exact solutions 
of Einstein's equations the solution is symmetric with respect to the 
equatorial plane if certain multipole moments of the mass-, angular momentum-, 
electric and magnetic field-distributions vanish \cite{MMR}. This is obtained 
if the five parameters are real \cite{MSM}. Because of the imaginary parameter 
$b$, terms of the multipole moments change their role from mass to angular 
momentum multipole moments or from electric to magnetic field multipole 
moments and vice versa, and destroy this symmetry. One has to test if the 
physical gravitational and gravitomagnetic forces are still symmetric. 
Using the formalism of the 3+1-split \cite{TPM} the four-dimensional 
space-time can be brought into the form
\begin{equation} \label{Splitmetrik}
ds^2 = \alpha^2\,dt^2 - h_{ik}\,(dx^{i} + \beta^{i}\,dt)\,(dx^{k} 
+ \beta^{k}\,dt) .
\end{equation}
To perform the following calculations it is convenient to introduce a 
``star-fixed'' coordinate system, at best in Boyer-Lindquist-like 
quasi-spherical coordinates. After comparison of (\ref{Splitmetrik}) with the 
metric (\ref{Metrik}) in these coordinates 
\begin{eqnarray}
ds^2 &=& f\,(dt - w \, d\phi)^2 - f^{-1}\,\Bigl[ \,e^{2 \gamma}\,\Bigl(
\frac{(r - m)^2 - \kappa^2\,\cos{\theta}^2}{(r - m)^2 - \kappa^2}\,dr^2 
\nonumber \\
&& + [ (r - m)^2 - \kappa^2\,\cos{\theta}^2 ]\,d\theta^2 \Bigr) \nonumber \\
&& + [ (r - m)^2 - \kappa^2 ]\,\sin{\theta}^2\,d\phi^2 \Bigr] 
\end{eqnarray}
-- for the transformation of the coordinates $\rho$ and $z$ to $r$ and 
$\theta$ see eq. (\ref{coord_transform}) -- it follows
\begin{eqnarray} \label{mankosplit}
\alpha^2 &=& \frac{f\,[ (r - m)^2 - \kappa^2 ]\,\sin{\theta}^2}{[ (r - m)^2 
- \kappa^2 ]\,\sin{\theta}^2 - f^2\,w^2} \nonumber \\
\beta^{\phi} &=& \frac{f^2\,w}{[ (r - m)^2 - \kappa^2 ]\,\sin{\theta}^2 - 
f^2\,w^2} \nonumber \\ 
\beta^{r} &=& 0 \qquad \beta^{\theta} = 0 \nonumber \\
h_{rr} &=& \frac{e^{2 \gamma}\,[ (r - m)^2 - \kappa^2\,\cos{\theta}^2 ]}{f\,
[ (r - m)^2 - \kappa^2 ]} \nonumber \\
h_{\theta\theta} &=& \frac{e^{2 \gamma}\,[ (r - m)^2 - \kappa^2\,
\cos{\theta}^2 ]}{f} \nonumber \\
h_{\phi\phi} &=& \frac{[ (r - m)^2 - \kappa^2 ]\,\sin{\theta}^2 - f^2\,w^2}{f} 
\nonumber \\
h_{ij} &=& 0 \qquad \,\,i \ne j .
\end{eqnarray}
The gravitational and gravitomagnetic force can be calculated with
\begin{equation} \label{Kraftpotentiale}
g_{i} = - \frac{1}{\alpha}\,\alpha_{|i} \qquad \qquad \qquad
H_{ij} = \frac{1}{\alpha}\,\beta_{j|i} ,
\end{equation}
which means
\begin{eqnarray} \label{grav_forces}
\bmath{g} &=& - \frac{1}{\alpha}\,\nabla\,\alpha = - \frac{1}{\alpha}\,\left( 
\frac{1}{\sqrt{h_{rr}}}\,\frac{\partial \alpha}{\partial r}\,
\bmath{e_{\hat r}} + \frac{1}{\sqrt{h_{\theta\theta}}}\,\frac{\partial 
\alpha}{\partial \theta}\,\bmath{e_{\hat \theta}} \right) \nonumber \\
H_{\hat \theta \hat \phi} &=& \frac{1}{\alpha}\,\left( \frac{1}{2}\,
\frac{1}{\sqrt{h_{\theta\theta}}}\,\frac{1}{\sqrt{h_{\phi\phi}}}\,\beta^{\phi}
\,\frac{\partial h_{\phi\phi}}{\partial \theta} \right. \nonumber \\
&& \left. + \frac{1}{\sqrt{h_{\theta
\theta}}}\,\frac{1}{\sqrt{h_{\phi\phi}}}\,h_{\phi\phi}\,
\frac{\partial \beta^{\phi}}{\partial \theta} \right) \nonumber \\
H_{\hat \phi \hat \theta} &=& \frac{1}{\alpha}\,\left( - \frac{1}{2}\,
\frac{1}{\sqrt{h_{\theta\theta}}}\,\frac{1}{\sqrt{h_{\phi\phi}}}\,\beta^{\phi}
\,\frac{\partial h_{\phi\phi}}{\partial \theta} \right) \nonumber \\
H_{\hat r \hat \phi} &=& \frac{1}{\alpha}\,\left( \frac{1}{2}\,
\frac{1}{\sqrt{h_{rr}}}\,\frac{1}{\sqrt{h_{\phi\phi}}}\,\beta^{\phi}
\,\frac{\partial h_{\phi\phi}}{\partial r} \right. \nonumber \\
&& \left. + \frac{1}{\sqrt{h_{rr}}}\,
\frac{1}{\sqrt{h_{\phi\phi}}}\,h_{\phi\phi}\,
\frac{\partial \beta^{\phi}}{\partial r} \right) \nonumber \\
H_{\hat \phi \hat r} &=& \frac{1}{\alpha}\,\left( - \frac{1}{2}\,
\frac{1}{\sqrt{h_{rr}}}\,\frac{1}{\sqrt{h_{\phi\phi}}}\,\beta^{\phi}
\,\frac{\partial h_{\phi\phi}}{\partial r} \right) \nonumber \\
\end{eqnarray}
with the orthonormal basis vectors
\begin{eqnarray}
\bmath{e_{\hat r}} &=& \frac{1}{\sqrt{h_{rr}}}\,\frac{\partial}{\partial r} 
\nonumber \\
\bmath{e_{\hat \theta}} &=& \frac{1}{\sqrt{h_{\theta\theta}}}\,
\frac{\partial}{\partial \theta} \nonumber \\
\bmath{e_{\hat \phi}} &=& \frac{1}{\sqrt{h_{\phi\phi}}}\,
\frac{\partial}{\partial \phi} .
\end{eqnarray}
Inserting all functions in (\ref{grav_forces}), one finds that the
 transformation $r, \theta = \frac{\pi}{2} - \vartheta \to r, \theta = 
\frac{\pi}{2} + \vartheta$ does not change the coefficient of 
$\bmath{e_{\hat r}}$ and the components $H_{\hat r \hat \phi}$ and $H_{\hat 
\phi \hat r}$, but changes the sign of the coefficient of $\bmath{e_{\hat 
\theta}}$ and the components $H_{\hat \theta \hat \phi}$ and $H_{\hat \phi 
\hat \theta}$. This shows that the physical forces are symmetric with respect 
to the equatorial plane.

\section{The ergoregion} \label{erg}

The existence of the ergoregion is a property of stationary, axisymmetric 
space-times which is defined as the region between an event horizon (if 
existing) and the stationary limit surface. 
In stationary, axisymmetric space-times the components $g_{\mu\nu}$ of the 
metric tensor are independent of the cyclic variables $t$ and $\phi$. Therefore
two Killing vectors 
\begin{equation}
\xi_{(t)}^{\mu} = \frac{\partial}{\partial t} \qquad \textrm{und}
\qquad \xi_{(\phi)}^{\mu} = \frac{\partial}{\partial \phi} .
\end{equation}
are existing. With the angular velocity
\begin{equation}
\Omega = \frac{d\phi}{dt} = \frac{d\phi/d\tau}{dt/d\tau} = \frac{u^{\phi}}{u^t}
\end{equation}
the four-velocity of a stationary observer takes the form
\begin{eqnarray} \label{Vierergeschw}
u^{\mu} &=& u^t (\frac{\partial}{\partial t} + \Omega
\frac{\partial}{\partial \phi}) = \frac{\xi_{(t)}^{\mu} + \Omega
\xi_{(\phi)}^{\mu}}{|\xi_{(t)}^{\mu} + \Omega \xi_{(\phi)}^{\mu}|} \nonumber \\
&=& \frac{\xi_{(t)}^{\mu} + \Omega \xi_{(\phi)}^{\mu}}{\sqrt{g_{tt} + 2
\Omega g_{t\phi} + \Omega^2 g_{\phi \phi}}} 
\end{eqnarray}
with which the condition for $\Omega$ 
\begin{equation} \label{Omega}
\Omega_{-} < \Omega < \Omega_{+} \quad
\Omega_{\pm} = \omega \pm \sqrt{\omega^2 - \frac{g_{tt}}{g_{\phi
\phi}}} \quad \omega = - \frac{g_{\phi
t}}{g_{\phi \phi}} 
\end{equation}
is introduced. There is a limit where no static observer with $\Omega = 0$ can 
exist. The surface of this stationary limit can be calculated from
\begin{equation} \label{ergos}
g_{tt} = 0 \qquad \Longrightarrow \qquad f = 0 .
\end{equation}

\begin{figure}
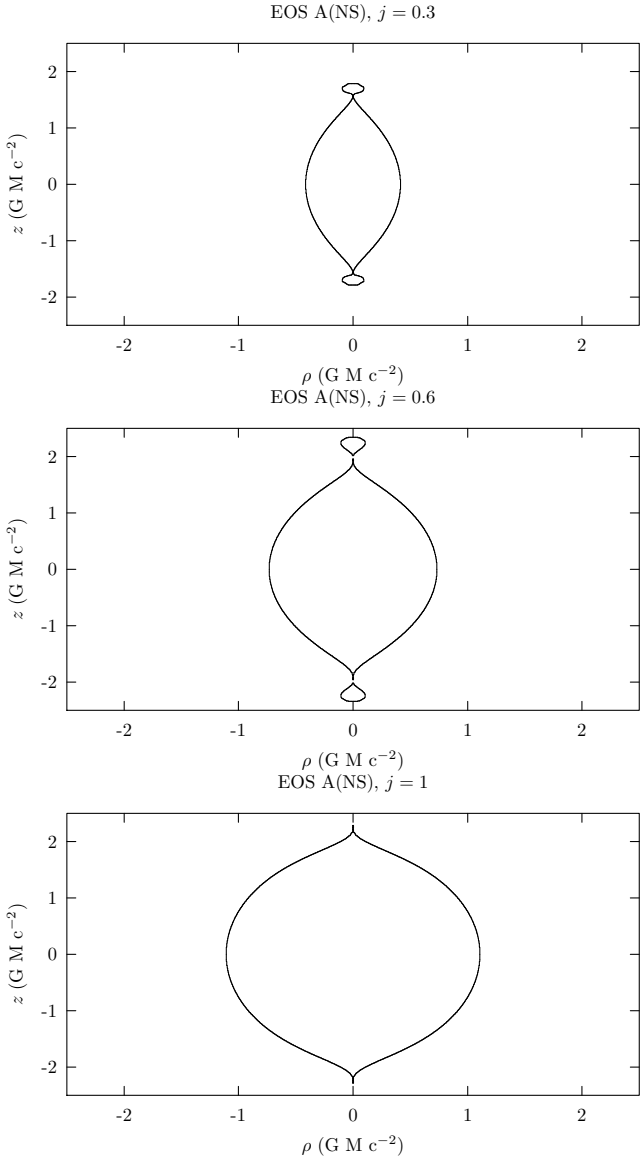

\resizebox{\hsize}{!}{\includegraphics{ergos_ANS_1.eps}}
\resizebox{\hsize}{!}{\includegraphics{ergos_ANS_2.eps}}
\resizebox{\hsize}{!}{\includegraphics{ergos_ANS_3.eps}}
\caption{Inflation of the ergoregion with increasing rotation parameter $j$}
\label{ergos_ANS}
\end{figure}

The implicit plots of this surface reveal new features which are unknown for 
the Kerr-solution. In contrast to the Kerr-metric, the ergoregion 
inflates with increasing rotation parameter $j$ for neutron stars in their 
normal sequence (their mass is below the maximum of a non-rotating star), 
while the EOS and the rest mass of the star are kept fixed 
(Fig. \ref{ergos_ANS}). 

Additionally the ergoregion consists beside the main body of two ``bubbles'' 
lying at the poles of the ergoregion whose existence is completely unknown in 
the Kerr-metric. As the rotation speed increases, the bottle neck connecting 
the bubbles with the main ergoregion attenuates until, at a critical 
$j^{peel}_{crit}$, these bubbles peel away.

\begin{figure}
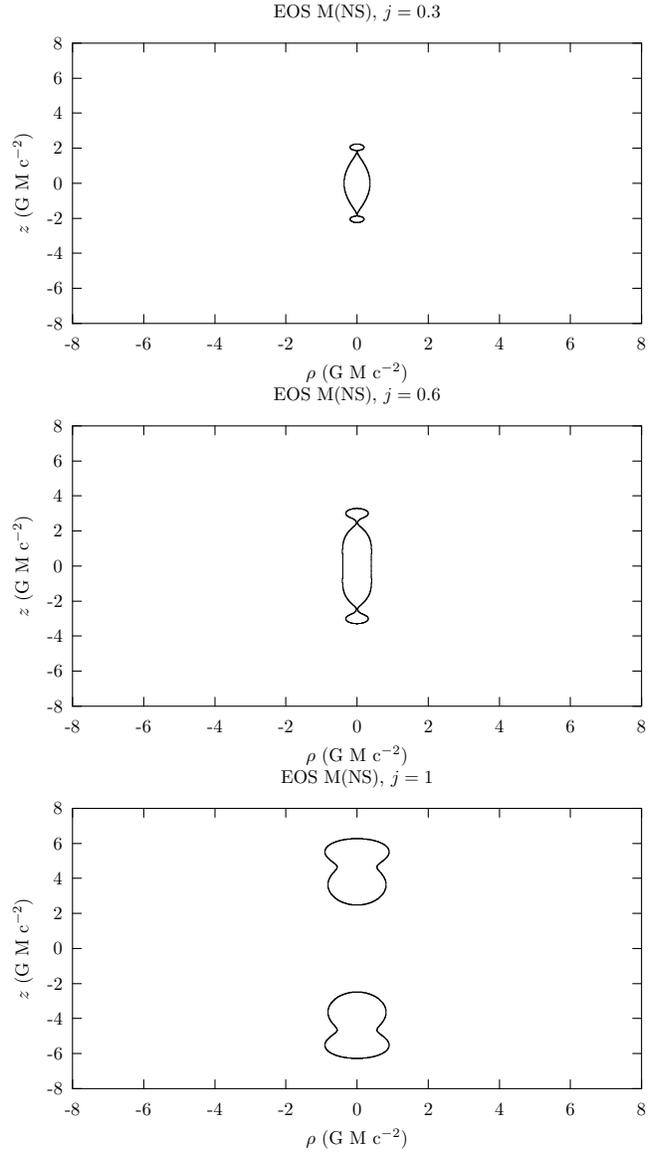

\resizebox{\hsize}{!}{\includegraphics{ergos_MNS_1.eps}}
\resizebox{\hsize}{!}{\includegraphics{ergos_MNS_2.eps}}
\resizebox{\hsize}{!}{\includegraphics{ergos_MNS_3.eps}}
\caption{Constriction of the ergoregion in the equatorial plane and splitting 
into two main ergoregions at a critical rotation}
\label{ergos_MNS}
\end{figure}

Another possible effect of rapid rotation is a constriction of the ergoregion 
in the equatorial plane which also increases for higher rotation. At 
another critical rotation value $j^{constr}_{crit}$ two separated regions 
appear (Fig. \ref{ergos_MNS}). This takes place only on neutron stars with  
stiffer EOS FPS, L and M in the normal sequence (notation see Cook et al. 
\shortcite{CST}).
 
\begin{figure}
\resizebox{\hsize}{!}{\includegraphics{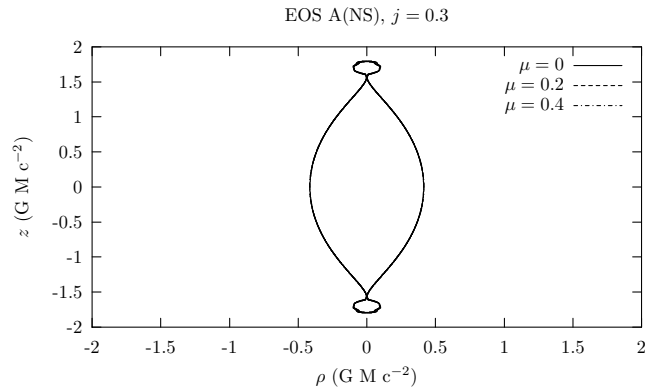}}
\caption{Invariance of the main ergoregion to an increasing magnetic dipole 
moment $\mu$}
\label{ergos_ANS_mu}
\end{figure}

\begin{figure}
\resizebox{\hsize}{!}{\includegraphics{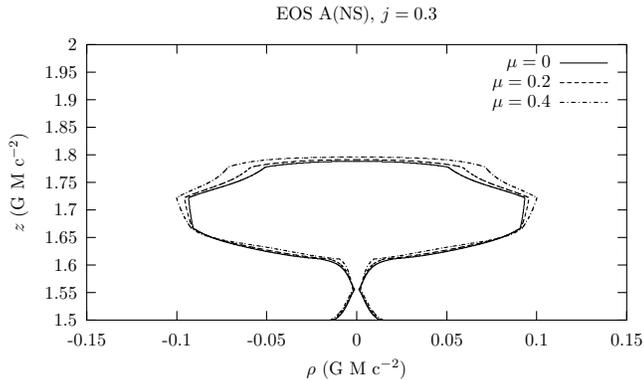}}
\caption{Effects of an increasing magnetic dipole moment $\mu$ to the polar 
bubbles}
\label{ergos_ANS_mu_detail}
\end{figure}

Until now, the magnetic dipole moment has been neglected. The surprising 
result of an increased non-vanishing magnetic field is the fact that the main 
ergoregion is almost not affected (Fig. \ref{ergos_ANS_mu}), while the bubbles 
grow in size and the constriction moves to larger values of $z$ coordinate  
(Fig. \ref{ergos_ANS_mu_detail}). Because of the fact that observed neutron 
stars have magnetic fields in the range of $10^{12}$ - $10^{15}$ gauss, the 
parameter $\mu$ should be smaller than $0.1$. Therefore, the effect of a 
non-vanishing magnetic field is relatively small.
Neutron stars are flattened due to their rotation and oblate; their equatorial 
radius is larger than the distance between their poles, whereas the ergoregion
is prolate. Three different cases are then possible:
\begin{enumerate}
\item $r^{erg}_{eq} > r^{*}_{eq}$, the ergoregion is completely outside the 
star
\item $r^{erg}_{eq} < r^{*}_{eq}$, but $r^{erg}_{pole} > r^{*}_{pole}$, the 
      ergoregion exceeds the stellar surface only in the polar region
\item $r^{erg}_{eq} < r^{*}_{eq}$ and $r^{erg}_{pole} < r^{*}_{pole}$, the 
      ergoregion is completely inside the star.
\end{enumerate}
Table \ref{erg_table} shows which case occurs for which EOS if $j$ is smaller 
than the $j$ at mass-shedding for uniform rotation.
\begin{table}
\caption{Position of the ergoregion in comparison to the stellar surface
($Q = 0$, $\mu = 0$)}
\label{erg_table}
\begin{tabular}{lll}
\hline
Equation of state & $j$ & case \\
\hline\hline
A normal sequence & 0.3 & (iii) \\
& 0.6 & (iii) \\
A maximum mass normal sequence & 0.3 & (ii) \\
& 0.6 & (ii) \\
A supramassive sequence & 0.6 & (ii) \\
\hline
AU normal sequence & 0.3 & (iii) \\
& 0.6 & (ii) \\
AU maximum mass normal sequence & 0.3 & (ii) \\
& 0.6 & (ii) \\
AU supramassive sequence & 0.6 & (ii) \\
\hline
FPS normal sequence & 0.3 & (iii) \\
& 0.6 & (iii) \\
FPS maximum mass normal sequence & 0.3 & (ii) \\
& 0.6 & (ii) \\
FPS supramassive sequence & 0.6 & (i) \\
\hline
L normal sequence & 0.3 & (iii) \\
& 0.6 & (iii) \\
L maximum mass normal sequence & 0.3 & (ii) \\
& 0.6 & (ii) \\
L supramassive sequence & 0.6 & (ii) \\
\hline
M normal sequence & 0.3 & (iii) \\
& 0.6 & (iii) \\
M maximum mass normal sequence & 0.3 & (iii) \\
& 0.6 & (iii) \\
M supramassive sequence & 0.6 & (ii) \\
\hline
\end{tabular}
\end{table}
For all EOS in the normal sequence the ergoregion is inside the region for 
slow rotation, where the stellar surface lies in the models of the interior. 
Increasing $j$, the ergoregion moves outside in the polar region 
after passing a critical $j^{erg}_{crit}$. Only stars with the EOS AU can be 
accelerated to this critical value, all other stars become unstable due to 
mass-shedding at smaller $j$. The critical $j$ decreases with increasing 
stellar mass and increases with increasing stiffness.

\section{Singularities of the global solution}

Singularities are given if the curvature tensor diverges. This 
divergence occurs if the denominator of the Ernst-potentials 
\cite{Ern} has to vanish \cite{MMS}:
\begin{equation} \label{singeq}
A + 2 m B = 0
\end{equation}
with 
\begin{eqnarray} \label{Metrik-Ernstpot_Abk}
A & = & 4 [ (\kappa^2 x^2 - \delta y^2)^2 - d^2 - i \kappa^3 x y (a - b)
(x^2 - 1)] \nonumber \\
&& - (1 - y^2) [(a - b) (d - \delta) - m^2 b + Q \mu]  \nonumber \\
&& [(a - b) (y^2 + 1) + 4 i \kappa x y] \nonumber \\
B & = & \kappa x \{2 \kappa^2 (x^2 - 1) + [b (a - b) + 2 \delta] (1 -
y^2)\} \nonumber \\
&& + i y \{ 2 \kappa^2 b (x^2 - 1) - [\kappa^2 (a - b) - m^2 b + Q \mu  
\nonumber \\
&& - 2 a \delta] (1 - y^2)\} .
\end{eqnarray}
Equation (\ref{singeq}) in fact consists of two equations, because $A$ and 
$B$ are complex and both, the real and imaginary part, have to vanish 
simultaneously. These two equations give the coordinates $x$-$y$, or 
$\rho$-$z$ respectively, of the singularities.

\begin{figure}
\resizebox{\hsize}{!}{\includegraphics{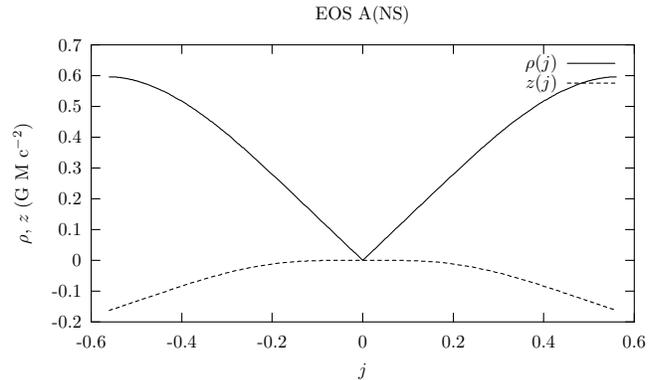}}
\caption{Splitting of the point singularity ($j = 0$) into two ring 
singularities ($j \neq 0$); only the singularity below the equatorial plane is
shown ($z < 0$) due to clarity (dashed line)}
\label{sing_ANS}
\end{figure}

\begin{figure}
\resizebox{\hsize}{!}{\includegraphics{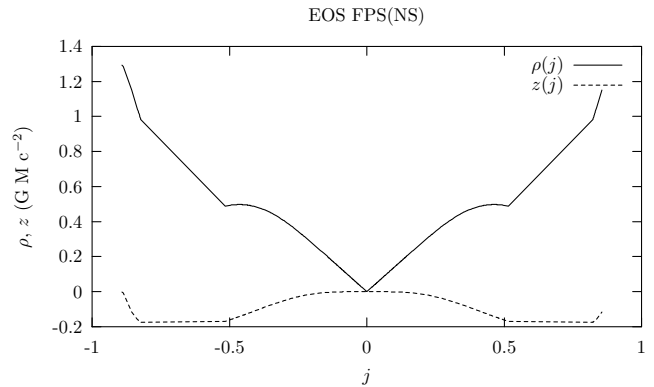}}
\caption{Same as Fig. \ref{sing_ANS}, but also merging of the two ring 
singularities to one single ring singularity ($\rho \neq 0$, $z = 0$) at 
$|j| \approx 0.9$}
\label{sing_FPSNS}
\end{figure}

\begin{figure}
\resizebox{\hsize}{!}{\includegraphics{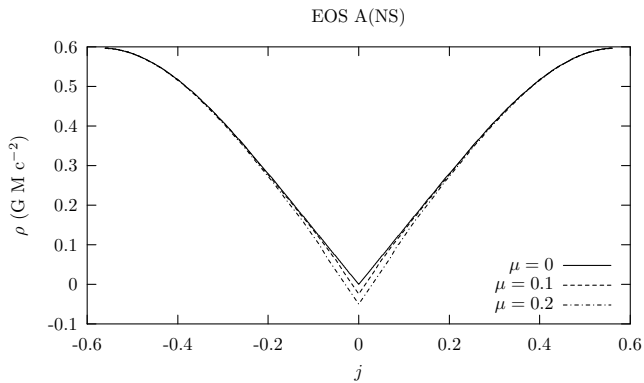}}
\caption{Radius of the ring singularities; two ring singularities for 
vanishing rotation ($\rho \neq 0$), merging at a critical rotation (sign 
change of $\rho$)}
\label{sing_ANS_mu_rho}
\end{figure}

\begin{figure}
\resizebox{\hsize}{!}{\includegraphics{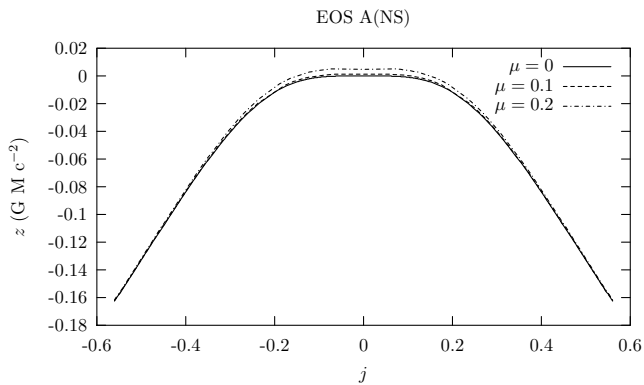}}
\caption{Distance of the ring singularities to the equatorial plane; moving of 
the singularities from above to below the equatorial plane (sign change of $z$)}
\label{sing_ANS_mu_z}
\end{figure}

These were computed numerically for all EOS and sequences as a
function of the rotation parameter $j$. If $j = 0$ only one point singularity 
arises in the origin as expected for the Schwarzschild limiting case. With 
increasing rotation this singularity splits into two ring singularities whose 
radius increases and which move away from the equatorial plane 
(Fig. \ref{sing_ANS}). In a few cases the singularities merge 
again to one single ring singularity at a critical $j^{sing}_{crit}$ 
(Fig. \ref{sing_FPSNS}), but the data of Cook et al. \shortcite{CST} suggest 
that stars become unstable before reaching this rapid rotation.
Again the results shown apply only to non-magnetized stars. In the presence of
magnetic fields the two ring singularities are present ($\rho \neq 0$, 
$z \neq 0$) even for vanishing rotation. The Schwarzschild space-time is of 
course no limiting case anymore. Increasing $j$, the singularities approach 
each other, merge at a critical rotation $j^{radius}_{crit}$ -- but not in the 
origin -- (sign change of $\rho$ while $z \neq 0$) and seem to change their 
roles (sign change of $z$) at $j^{dist}_{crit}$. The singularity lying above 
the equatorial plane for slow rotation moves after the merging below the 
equatorial plane (Fig. \ref{sing_ANS_mu_rho}, \ref{sing_ANS_mu_z}).

\section{Horizons}

The horizon is the null surface spanned by the tangential vectors $\partial_{t}$
and $\partial_{\phi}$. The resulting condition is
\begin{equation} \label{horizon}
g_{tt}\,g_{\phi\phi} - g_{\phi\,t}^{2} = - \rho^{2} = 0 .
\end{equation}
$\rho$ plays in this coordinate system the role of the Weyl radius.

The singularities are located at $\rho > 0$. Therefore the solution hurts the 
``cosmic censorship hypothesis''. But because of the fact that all 
singularities are within the stellar surface this is irrelevant.

\section{Conclusions}

The same procedure as shown in section \ref{fit} was already performed -- 
in the opposite order -- by Sibgatullin \& Sunyaev \shortcite{SiS1}, but only 
with the normal sequences of EOS A, EOS AU and EOS FPS and with an older 
analytic solution found by Manko et al. \shortcite{MMR} which did not permit a 
rational function representation and could not be written in a concise form. 
These three EOS are of lower stiffness. These authors arrived at the 
conclusion that it is sufficient to include only the mass quadrupole moment 
and no higher multipole moments -- a result we could not confirm for all, 
especially not for the stiffest EOS L and M. The fact that the accuracy is 
high in the case of the redshift, but much poorer in describing the radius of 
the marginally-stable-orbit which is more sensitive to higher multipole 
moments of the space-time (see comparison to Kerr in the figures), seems to 
show that higher multipole moments have to be included, especially for rapid 
rotation and stiffer EOS, to improve the form of the exterior gravitational 
field of neutron stars. But until a new exterior solution is found, this new 
self-consistent relativistic model of rapidly rotating neutron stars, we 
present here, is a great improvement in comparison to the Schwarzschild- and 
Kerr-solutions sometimes used to describe also rapidly rotating stars.

\begin{figure}
\resizebox{\hsize}{!}{\includegraphics{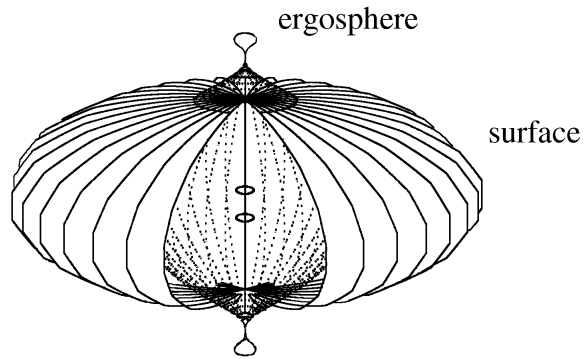}}
\caption{Ring currents as sources of the exterior fields}
\label{ringcurrent}
\end{figure}
It is clear that a vacuum solution is not able to describe the interior of 
the star. For each EOS the parameters have to be adjusted to 
match both regions. Therefore this solution cannot claim to be global and 
the presence of (naked) singularities which are always inside the stellar 
surface is irrelevant for astrophysics. However they help to interprete the 
exterior field as a space-time created by a configuration with two ring 
currents (Fig. \ref{ringcurrent}). These mass- and angular momentum currents 
arise as singularities in the curvature tensor. Coupled by the Einstein 
equations to the energy-momentum-tensor of the electromagnetic fields these 
sources should also appear in the electric and magnetic field-distributions. 
Because of charge neutrality the ring currents must have different rotation 
directions and therefore create also weak electromagnetic quadrupole fields.
As discussed in section \ref{erg} it seems to be common that the ergoregion is 
not completely enclosed by the stellar surface. The fact that it leak from 
the interior of the star only in the polar regions could have very interesting 
consequences for the acceleration of particles and propagation of light.

\end{document}